\documentclass[a4paper,11pt]{article}
\usepackage{pos}
\usepackage{placeins}
\usepackage{multirow}
\usepackage{braket}

\usepackage[font=small, labelfont=bf, textfont={small,it}]{caption}
\title{Scale setting of $\mathrm{SU}(N)$ Yang--Mills theories via Twisted Gradient Flow}
\ShortTitle{SU(N) scale setting via TGF}

\author[a]{Claudio Bonanno}
\author[b]{Jorge Luis Dasilva Gol\'an}
\author[c]{Massimo D'Elia}
\author[a]{Margarita Garc\'ia P\'erez}
\author*[c]{Andrea Giorgieri}

\affiliation[a]{Instituto de F\'isica T\'eorica UAM-CSIC, Calle Nicol\'as Cabrera 13-15,\\Universidad Aut\'onoma de Madrid, Cantoblanco, E-28049 Madrid, Spain}

\affiliation[b]{Brookhaven National Laboratory (BNL), PO Box 5000, Upton, NY 11973-5000, United States}

\affiliation[c]{Dipartimento di Fisica dell'Universit\`a di Pisa \& \\ INFN Sezione di Pisa, Largo Pontecorvo 3, I-56127 Pisa, Italy\\\vspace{\baselineskip}}

\emailAdd{claudio.bonanno@csic.es}
\emailAdd{jgolandas@bnl.gov}
\emailAdd{massimo.delia@unipi.it}
\emailAdd{margarita.garcia@uam.es}
\emailAdd{andrea.giorgieri@phd.unipi.it}

\abstract{
We present preliminary results for the scale setting of $\mathrm{SU}(N)$ Yang--Mills theories using twisted boundary conditions and the gradient-flow scale $\sqrt{t_0}$. The end goal of this study is to determine the $\mathrm{SU(N)}$ $\Lambda$-parameter through the \emph{step-scaling} method. The scale $\sqrt{t_0}$, being defined from the flowed action density of the gauge fields, is correlated with their topological charge and thus could be affected by \emph{topological freezing}. We deal with this problem with the Parallel Tempering on Boundary Conditions algorithm, which we found to be effective for the same numerical setup in a previous work.
}

\FullConference{The 41st International Symposium on Lattice Field Theory (LATTICE2024)\\
	28 July - 3 August 2024\\
	Liverpool, UK\\}

\newcommand{\be}{\begin{equation}}
\newcommand{\ee}{\end{equation}}
\newcommand{\beq}{\begin{eqnarray}}
\newcommand{\eeq}{\end{eqnarray}}
\newcommand{\beqnn}{\begin{eqnarray*}}
\newcommand{\eeqnn}{\end{eqnarray*}}
\newcommand{\SU}{\mathrm{SU}}

\newcommand{\ov}{\mathrm{ov}}
\newcommand{\clov}{\mathrm{clov}}

\newcommand{\Tr}{\mathrm{Tr}}

\newcommand{\tl}{\tilde l}
\newcommand{\tL}{\tilde L}
\newcommand{\had}{\mathrm{had}}

\graphicspath{{./}}

\begin{document}
\maketitle

\section{Introduction}

Our goal is to determine the $\Lambda$-parameter of $\SU(N)$ Yang--Mills theories using the renormalization scheme known as \emph{Twisted Gradient Flow} (TGF)~\cite{Ramos:2014kla,Bribian:2019ybc,Bribian:2021cmg}. This calculation can be divided into two steps. First, $\Lambda$ can be determined in units of a low energy renormalization scale $\mu_\had$ through the \emph{step-scaling} technique~\cite{Luscher:1991wu}, which consists in flowing the renormalization group from IR to UV scales in discrete steps to match lattice calculations with perturbation theory. Then, one can set the scale of the theory and determine $\mu_\had\sqrt{8t_0}$, where $t_0$ is a conventional reference scale defined via the gradient flow, so that also $\Lambda$ can be expressed in units of $t_0$. The value of $\Lambda/\mu_\had$ for $N=3,5,8$ has been determined in Ref.~\cite{DasilvaGolan:2023yvg}. Here we present preliminary results of the scale setting for $N=5$.

The determination of $t_0$ can be biased by \emph{topological freezing}, a well-known problem of standard algorithms in the sampling of the topological modes of Yang--Mills theories close to the continuum limit~\cite{Alles:1996vn,DelDebbio:2004xh,Schaefer:2010hu}. For this reason, we employ an algorithm specifically designed to mitigate topological freezing, the \emph{Parallel Tempering on Boundary Conditions} (PTBC)~\cite{Hasenbusch:2017unr,Bonanno:2020hht,Bonanno:2024zyn}. The PTBC also allows us to evaluate the possible bias on the scale setting of a frozen algorithm.

This manuscript is organized as follows: in Sec.~\ref{sec:scale} we explain the scale setting procedure and the effect of topology, in Sec.~\ref{sec:alg} we describe our numerical setup and the PTBC algorithm, in Sec.~\ref{sec:res} we present our preliminary results, and finally in Sec.~\ref{sec:conclu} we draw our conclusions.

\section{Scale setting and the effect of topology}\label{sec:scale}

The scale of $\SU(N)$ Yang--Mills theories can be conveniently set using the gradient flow~\cite{Narayanan:2006rf,Lohmayer:2011si,Luscher:2009eq}, a smoothing procedure that evolves the gauge fields $A_\mu(x)$ in a time $t$ according to the flow equation
\be\label{eq:wilson_flow}
\partial_t B_\mu (x, t) = D_\nu F_{\nu \mu} (x, t), \quad B_\mu (x, t = 0) = A_\mu (x)\, ,
\ee
where $D_\mu$ and $ F_{\mu \nu}$ are the covariant derivative and the field strength tensor of the flowed fields $B_\mu(x, t)$. The gradient-flow scale $t_0$ is defined for $\SU(3)$ as~\cite{Luscher:2010iy}:
\be\label{eq:t0_su3}
\left.\langle t^2E(t)\rangle\right|_{t=t_0} = 0.3\, ,
\ee
where $E(t)$ is the energy density of the flowed gauge fields,
\be\label{eq:clover_density}
E(t) = \frac{1}{2} \Tr \left [F_{\mu \nu} (x, t)F_{\mu \nu} (x, t)\right]\, .
\ee
In physical units this corresponds to $\sqrt{8t_0}\simeq 0.5 \text{ fm}$. A possible generalization to $\SU(N)$ is
\be\label{eq:t0_sun}
\frac{N}{N^2-1}\left.\langle t^2E(t)\rangle\right|_{t=t_0} = 0.1125 \, .
\ee
This definition coincides with Eq.~\eqref{eq:t0_su3} for $N=3$ and is normalized to cancel the $N$-dependence of the leading-order term of the small-t perturbative expansion of $E(t)$~\cite{Ce:2016awn}. The determination of $\mu_\had\sqrt{8t_0}$, in combination with the result for $\Lambda/\mu_\had$, allows to obtain $\Lambda\sqrt{8t_0}$. 

The determination of $t_0$ can be biased by topological freezing. To understand why, one should consider that the flowed energy density of a field configuration is correlated with its topological charge $Q$: the gradient flow drives the configuration towards a minimum of the action in its topological sector and this minimum increases with $|Q|$. Thus, a bias in the sampling of topology can affect $E(t)$ and so also $t_0$. In particular, the average energy density of the zero topological sector, in which algorithms are usually frozen, is expected to have power-like finite-volume corrections, while the volume dependence is exponentially suppressed if all topological sectors are considered~\cite{Brower:2003yx}. To evaluate this effect, we also consider a scale $t_0^{(0)}$ defined in the zero topological sector,
\be\label{eq:t0_sun_proj}
\frac{N}{N^2-1}\left.\frac{\langle t^2E(t)\delta_{Q,0}\rangle}{\langle \delta_{Q,0}\rangle}\right|_{t=t_0^{(0)}} = 0.1125 \, ,
\ee
where $\delta_{Q,0}$ is a $\delta$-function restricting the calculation to gauge configurations with $Q=0$.

\section{Numerical setup}\label{sec:alg}

We employ the Twisted Gradient Flow scheme described in Ref.~\cite{Bonanno:2024nba}. Briefly, we discretize the pure-gauge $\SU(N)$ theory using the Wilson plaquette action on a $L^2 \times \tL^2$ lattice with $\tL = L/N$. We impose Twisted Boundary Conditions (TBCs)~\cite{tHooft:1979rtg,Gonzalez-Arroyo:1982hyq} along the short directions $\mu = 1,2$ and Periodic Boundary Conditions (PBCs) along $\mu = 0,3$. The lattice action is
\be\label{eq:lattice_action_TBC}
S_{\rm W}[U] = -\frac{\beta}{N}\sum_{x,\mu>\nu} Z_{\mu\nu}^*(x)\Re\Tr \left[ P_{\mu\nu}(x)\right],
\ee
where $\beta = 2N/g^2$ is the inverse bare coupling and $P_{\mu\nu}(x)$ is the plaquette,
\be
P_{\mu\nu}(x) = U_\mu(x) U_\nu(x+a\hat{\mu}) U_\mu^{\dag}(x+a\hat{\nu}) U^{\dag}_\nu(x) \, .
\ee
The factor $Z_{\mu\nu}(x)$ implements TBCs:
\be
Z_{\mu\nu}(x) = Z_{\nu\mu}^*(x) =
\begin{cases}
	e^{i 2 \pi k /N} & \text{if } (\mu,\nu)=(1,2) \text{ and } x_{\mu}=x_\nu=0 \,,\\
	1               & \text{otherwise.}
\end{cases}
\ee
The value of $k$, an integer coprime with $N$, can be chosen as part of the scheme. To avoid the appearance of tachyonic instabilities, the best way to approach the large-$N$ limit is to take $k$ and $N$ two steps apart in the Fibonacci sequence~\cite{Chamizo:2016msz}, that is $k=1,2,3$ for $N=3,5,8$ respectively.

As dimensionless energy density on the lattice, we use the clover-discretized definition
\be
E_{\clov}(t) = \frac{1}{2} \Tr\left[C_{\mu\nu}(x,t)C_{\mu\nu}(x,t)\right] \, ,
\ee
where $C_{\mu\nu}(x,t)$ is the clover operator on the $(\mu,\nu)$ plane in the site $x$ evaluated after the gauge links have been evolved for a flow time $t$. The gradient-flow is also used to define the topological charge on the lattice. Given the clover discretization
\be
Q_{\clov}(t) = \frac{1}{32\pi^2}\sum_{x,\mu\nu\rho\sigma}\varepsilon_{\mu\nu\rho\sigma}\Tr\left[C_{\mu\nu}(x,t)C_{\rho\sigma}(x,t)\right] \, ,
\ee
we define the physical topological charge as
\be
Q = Q_{\clov}\left(\sqrt{8t}=cl\right) \quad (c = 0.3) \, ,
\ee
where $l=aL$ is the physical extent of the lattice and $\sqrt{8t}$ is the \emph{smoothing radius} of the gradient flow. Analogously to the $\SU(3)$ case analyzed in Ref.~\cite{Bonanno:2024nba}, also for $N=5$ we verified that, with this choice of the flow time, $Q_{\clov}(t)$ has already reached a plateau in $t$ and is close to an integer number. Thus, we can define the lattice $\delta$-function to project on the zero topological sector in Eq.~\eqref{eq:t0_sun_proj} as
\be
\hat \delta(Q) =
\begin{cases}
1 & \text{if } \vert Q \vert < 0.5\\
0 & \text{otherwise}.
\end{cases}
\ee

In order to address topological freezing, we adopt the Parallel Tempering on Boundary Conditions (PTBC) algorithm of Ref.~\cite{Bonanno:2024nba}. We consider $N_r$ replicas $r=0,1,\dots,N_r-1$ of the lattice, each one differing for the boundary conditions imposed on a small sub-region called the \emph{defect} $D$. We choose $D$ to be an $L_d \times L_d \times L_d$ spatial cube, placed on the time boundary $x_0 = L-1$. Links that cross $D$ orthogonally (i.e., temporal links) are multiplied by a real factor $c(r)$. For the physical replica (i.e., the one on which observables are computed) $c(0)=1$, so the defect has no effect and links enjoy PBCs. The other replicas interpolate between periodic and open boundary conditions on the defect: $c(N_r-1)=0$ for the last replica and $0 < c(r) < 1$ for those in-between. The defect is implemented by taking as the action of the replica $r$
\begin{equation}
	S_{\rm W}^{\left(c(r)\right)}[U_r] = -\frac{\beta}{N}\sum_{x,\mu>\nu} K_{\mu\nu}^{\left(c(r)\right)}(x) Z_{\mu\nu}^*(x)\Re\Tr \left[ P^{(r)}_{\mu\nu}(x)\right] \, ,
\end{equation}
where $U_r$ denotes the gauge links of the replica $r$. The factor $K_{\mu\nu}^{\left(c(r)\right)}(x)$ changes the boundary conditions on the defect, similarly to the twist factor $Z_{\mu\nu}(x)$:
\be
	K_{\mu\nu}^{(c(r))}(x) = K_{\mu}^{(c(r))}(x)\,K_{\nu}^{(c(r))}(x+a\hat{\mu})\,K_\mu^{(c(r))}(x+a\hat{\nu})\,K_\nu^{(c(r))}(x) \, ,
\ee
\be
K_{\mu}^{\left(c(r)\right)}(x) = 
\begin{cases}
	c(r)  & \text{if } \mu=0\,,\,\, x_0=L-1\,, \text{ and } 0 \le x_1,x_2,x_3 < L_d\,,\\
	1     & \text{otherwise.}
\end{cases}
\ee

For what concerns the Monte Carlo sampling, each replica is updated simultaneously and independently performing 1 lattice sweep of the standard local heat-bath algorithm~\cite{Creutz:1980zw,Kennedy:1985nu}, followed by $n_{\ov}=12$ lattice sweeps of the standard local over-relaxation algorithm~\cite{Creutz:1987xi}. Then, swaps among two adjacent replicas $(r,s=r+1)$ are proposed and accepted via a Metropolis step with probability
\be
p(r,s) = \min\left\{1, e^{-\Delta S^{(r,s)}_{\rm swap}}\right\} \, ,
\ee
\be
    \Delta S^{(r,s)}_{\rm swap} = S_{\rm W}^{\left(c(r)\right)}[U_s]+S_{\rm W}^{\left(c(s)\right)}[U_r]-S_{\rm W}^{\left(c(r)\right)}[U_r]-S_{\rm W}^{\left(c(s)\right)}[U_s] \, .
\ee
The values $c(r)$ of intermediate replicas are tuned with short test simulations in order to achieve a mean acceptance of swaps around $20 \%$ for each pair of replicas. Thus, a given field configuration performs a sort of random walk among different replicas. Moreover, to improve the performance of the algorithm, the defect is translated randomly around the lattice and local updates are more frequent around it. 

\begin{figure}[!t]
\centering
\includegraphics[width=0.8\textwidth]{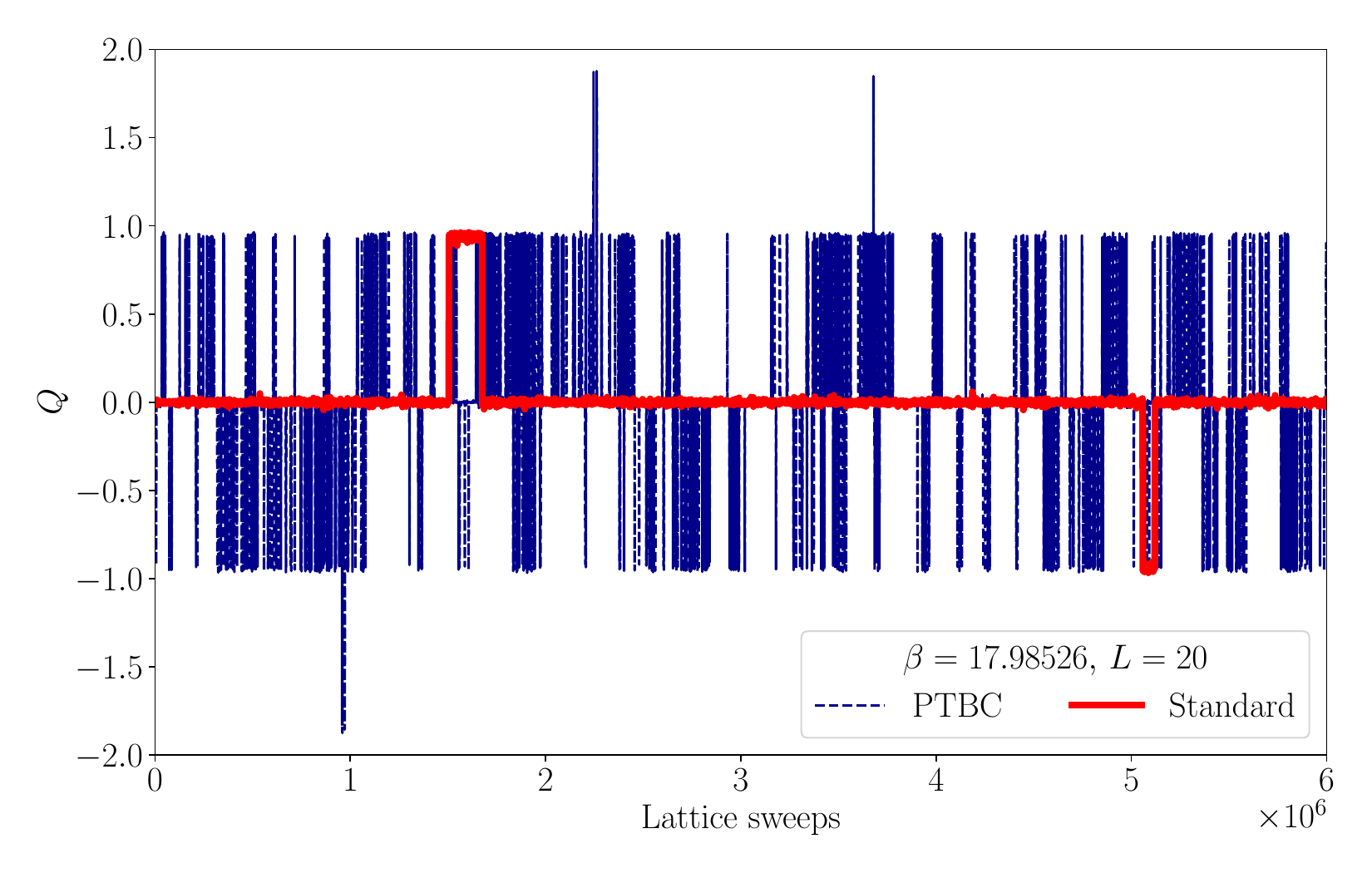}
\caption{Comparison of the Monte Carlo evolutions of the lattice topological charge $Q$ obtained with the PTBC and a standard algorithm in an $\SU(5)$ simulation. Only a fraction of the total statistics is shown. The PTBC uses $N_r=13$ replicas and the standard algorithm consists in the simulation of only the physical replica with the same combination of one heat-bath sweep followed by $n_\ov=12$ over-relaxation sweeps. The Monte Carlo time is expressed in units of total lattice sweeps, keeping into account all the replicas used in the PTBC for a fair comparison. The decorrelation of $Q$ achieved with the PTBC results in a significant reduction of the statistical uncertainty on $\langle Q^2 \rangle$. From simulations of comparable computational effort, the PTBC and the standard algorithm give $\langle Q^2 \rangle = 0.084(2)$ and $\langle Q^2 \rangle = 0.08(2)$ respectively. The algorithmic improvement of the PTBC can be quantified by the integrated auto-correlation time $\tau$ of $Q^2$, also expressed in units of total lattice sweeps. For the PTBC $\tau=2.5(3)\cdot10^2$, while for the standard algorithm $\tau\gtrsim10^5$. }
\label{fig:ptbc_std_comp}
\end{figure}

In Ref.~\cite{Bonanno:2024nba} we determined that the PTBC can efficiently reduce the auto-correlation times of topological charge in our numerical setup. An example is shown in Fig.~\ref{fig:ptbc_std_comp}, where we compare the Monte Carlo evolutions of the lattice topological charge obtained with the PTBC and a standard local algorithm.

\section{Results}\label{sec:res}

\begin{figure}[!t]
\centering
\includegraphics[width=0.8\textwidth]{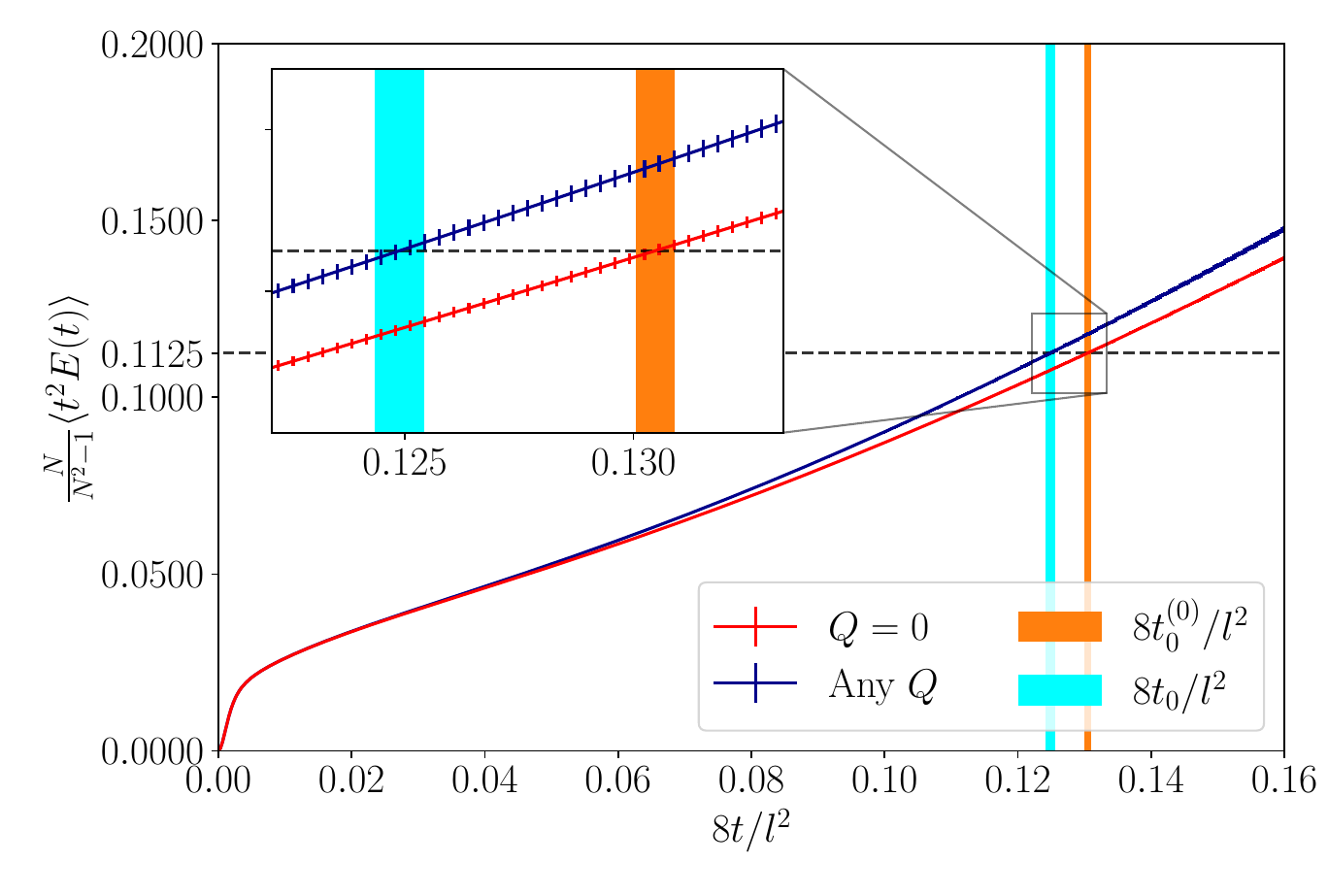}
\caption{Determination of $t_0$ in the $\SU(5)$ theory from the flow of the energy density of a lattice with $L=30,\ \beta=18.75186$. The scale $t_0^{\left(0\right)}$, defined by considering only gauge configurations with $Q=0$, is used to evaluate the effect of topological freezing on a standard algorithm.}
\label{fig:flow_comp}
\end{figure}

In this section, we present preliminary results of the scale setting for $\SU(5)$ and discuss the effects of topology and finite volumes. An example of the determination of $t_0$ is shown in Fig.~\ref{fig:flow_comp}. For each sampled gauge configuration, the flow equation Eq.~\eqref{eq:wilson_flow} is discretized and integrated with the adaptive third-order Runge-Kutta method described in Ref.~\cite{Fritzsch:2013je}. Then, the flow of the energy density $E(t)$ is interpolated to determine $t_0$ as defined in Eq.~\eqref{eq:t0_sun}. The modified scale $t_0^{(0)}$, defined in Eq.~\eqref{eq:t0_sun_proj}, is calculated considering only configurations with lattice topological charge $Q=0$ in the same ensemble. On the lattice, if the volume is large enough, we observe 
\begin{equation}
\langle t^2 E(t)\delta_{Q,0} \rangle/\langle \delta_{Q,0} \rangle \leq \langle t^2 E(t) \rangle\, .
\end{equation}
Since the threshold that defines the scale is approached from below, the projection to $Q=0$ results in a larger scale, $t_0^{(0)} \geq t_0$. The difference between the two definitions is the bias which can be expected from a standard algorithm suffering for topological freezing. However, the two definitions should converge in the infinite-volume limit~\cite{Brower:2003yx}.

A summary of the simulation points and the results of the scale setting is reported in Tab.~\ref{tab:simulation_points}. The renormalization scale $\mu_{\had}$ is defined as $a\mu_{\had} = 1/(0.3L_{\mu})$ (See Ref.~\cite{Bonanno:2024nba} for the details of the TGF scheme). For each one of the three lattice spacings, we simulated three volumes with $L\geq L_{\mu} \simeq 3\sqrt{8t_0}$.

\begin{table}[!t]
\begin{center}
\begin{tabular}{|c|c|c|c|c|c|c|}
\hline
\rule{0pt}{14pt}$\beta$ & $L_{\mu}$ & $N_r$ & $L_d$ & $L$ & $t_0/a^2$ & $t_0^{(0)}/a^2$ \\[2pt]
\hline
\multirow{3}{*}{17.98526} & \multirow{3}{*}{20} & \multirow{3}{*}{21} & \multirow{3}{*}{3} & 20 & 6.282(13) & 6.568(10) \\
& & & & 30 & 6.3509(61) & 6.5158(96) \\
& & & & 40 & 6.3967(64) & 6.416(12) \\
\hline
\multirow{3}{*}{18.75186} & \multirow{3}{*}{30} & \multirow{3}{*}{32} & \multirow{3}{*}{4} &  30 & 14.019(60) & 14.608(58) \\
& & & & 40 & 13.993(42) & 14.503(48) \\
& & & & 50 & 14.109(38) & 14.311(62) \\
\hline
\multirow{3}{*}{19.34158} & \multirow{3}{*}{40} & \multirow{3}{*}{44} & \multirow{3}{*}{5} & 40 & 25.349(82) & 26.21(10) \\
& & & & 50 & 25.366(68) & 25.762(89) \\
& & & & 60 & 25.302(85) & 25.631(81) \\
\hline
\end{tabular}
\end{center}
\caption{Summary of the simulation points of $\SU(5)$ and the results obtained for the gradient-flow scale $t_0$ and its modified version $t_0^{(0)}$ defined in the zero topological sector. $L_{\mu}$ defines the renormalization scale $a\mu_{\had} = 1/(0.3L_{\mu})$ in the TGF scheme. As for the PTBC parameters, the defect size $L_d$ is kept approximately constant in physical units and the number of replicas $N_r$ is tuned to have a $20\%$ acceptance of the swaps among replicas, as explained in Sec.~\ref{sec:alg}.}
\label{tab:simulation_points}
\end{table}

\begin{figure}[!t]
\centering
\includegraphics[width=0.8\textwidth]{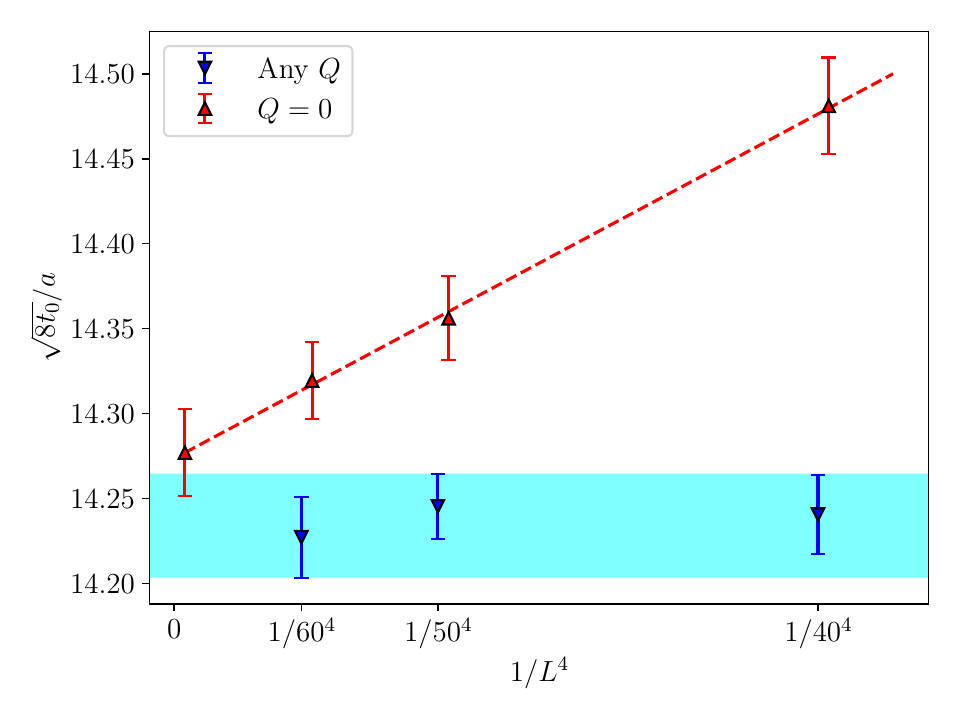}
\caption{Infinite-volume extrapolation of the $\SU(5)$ scales $t_0$ and $t_0^{\left(0\right)}$ determined at $\beta=19.34158$. The scale $\sqrt{8t_0}$ has no finite-volume effects within a $0.2\%$ uncertainty. The scale defined in the $Q=0$ sector shows a correction scaling as the inverse of the volume, but the infinite-volume extrapolation is compatible with the former.}
\label{fig:thermodynamic_limit}
\end{figure}

As expected, $t_0$ and $t_0^{(0)}$ seem to approach the same value in the infinite-volume limit, with $t_0^{(0)}$ showing larger finite-volume effects. As an example, let us discuss the data at the finest lattice spacing, shown in Fig.~\ref{fig:thermodynamic_limit}. In this case, all determinations of $t_0$ are compatible within a $0.2\%$ accuracy, while $t_0^{(0)}$ shows a significant volume dependence. However, the infinite-volume extrapolation of $t_0^{(0)}$ is compatible with $t_0$.

\FloatBarrier

\section{Conclusions}\label{sec:conclu}

We presented a preliminary study on the scale setting of the $\SU(5)$ Yang--Mills theory in the TGF scheme. This was done to be able to determine the $\SU(5)$ $\Lambda$-parameter in units of the gradient-flow scale $\sqrt{t_0}$ through the step-scaling method. We used the PTBC algorithm to mitigate topological freezing, which can introduce a bias in the scale setting.

Our preliminary analysis shows that, in the presence of a completely frozen topology, $t_0$ receives a positive bias with respect to the actual value obtained with a properly sampled topological charge. This bias seems to drop out on large volumes, as expected from general theoretical arguments. The present investigation will be further expanded in a forthcoming publication.

\acknowledgments
This work is partially supported by the Spanish Research Agency (Agencia Estatal de Investigación) through the grant IFT Centro de Excelencia Severo Ochoa CEX2020-001007-S and, partially, by the grant PID2021-127526NB-I00, both of which are funded by MCIN/AEI/10.13039/\linebreak501100011033. This work has also been partially supported by the project ”Non-perturbative aspects of fundamental interactions, in the Standard Model and beyond” funded by MUR, Progetti di Ricerca di Rilevante Interesse Nazionale (PRIN), Bando 2022, grant 2022TJFCYB (CUP I53D23001440006). We also acknowledge partial support from the project H2020-MSCAITN-2018-813942 (EuroPLEx) and the EU Horizon 2020 research and innovation programme, STRONG-2020 project, under grant agreement No. 824093. Numerical calculations have been performed on the \texttt{Leonardo} machine at Cineca, based on the agreement between INFN and Cineca, under the project INF24\_npqcd.

\bibliographystyle{JHEP}
\bibliography{biblio}

\end{document}